\documentclass[aps,prl%
,twocolumn%
,showpacs%
,superscriptaddress%
]{revtex4}

\usepackage{graphicx}
\usepackage{times}

\newcommand{\be}{\begin{equation}}
\newcommand{\ee}{\end{equation}}
\newcommand{\bea}{\begin{eqnarray}}
\newcommand{\eea}{\end{eqnarray}}

\newcommand{\affila}{
 Institut Curie, UMR 168,
 26 rue d'Ulm, F-75248 Paris C\'edex 05, France
}
\newcommand{\affilb}{ 
 Dept. of Biological Physics, E\"otv\"os University,
 P\'azm\'any P. stny. 1A, H-1117 Budapest, Hungary
}
\newcommand{\affilc}{ 
 Max Planck Institut f\"ur Physik komplexer Systeme,
 N\"othnitzer Str. 38, D-01187 Dresden, Germany
}

\begin{document}

\title{
 Formation and Interaction of Membrane Tubes
}

\author{Imre Der\'enyi}
 \affiliation{\affila}
 \affiliation{\affilb}
\author{Frank J\"ulicher}
 \affiliation{\affila}
 \affiliation{\affilc}
\author{Jacques Prost}
 \affiliation{\affila}

\date[]{submitted to Phys.\ Rev.\ Lett., February 8, 2002}

\begin{abstract}
We show that the formation of membrane tubes (or membrane
tethers), which is a crucial step in many biological processes, is
highly non-trivial and involves first order shape transitions. The
force exerted by an emerging tube is a non-monotonic function of its
length. We point out that tubes attract each other, which eventually
leads to their coalescence. We also show that detached tubes behave
like semiflexible filaments with a rather short persistence length.
We suggest that these properties play an important role in the
formation and structure of tubular organelles.
\end{abstract}


\maketitle

Biological membranes (such as the endoplasmic reticulum, the Golgi
apparatus, the inner mitochondrial membrane, or the plasma membrane)
often form highly dynamic tubular networks
\cite{tube}.
The formation and transport of membrane tubes (with tens of nanometers
in diameter) are thought to involve motor proteins that are able to
grab the membrane and pull on it as they move along the filaments of
the cytoskeleton
\cite{motion}.
Nanotubes (or tethers) can also be pulled out by various experimental
techniques (such as hydrodynamic flow
\cite{waugh82},
micropipettes
\cite{evans96},
or optical tweezers
\cite{raucher99}),
and very recently Roux {\it et al.}\
\cite{roux02}
have managed to set up a minimal {\it in vitro} experimental system, in
which tubes have been pulled by kinesin motor proteins. The physics of long
tubular membranes is relatively simple and is well understood. However,
the initial formation of nanotubes from planar or large spherical
pieces of membrane is a subtle process, the understanding of which is
crucial to the study of various biological processes (involving
tube formation or simply membrane pulling). In addition to tube
formation we also study the interaction of membrane tubes pulled from
the same membrane, and make a few comments on the mechanical
properties of detached membrane tubes.

Living cells maintain the surface tension of most of their membranes at
a constant level by keeping lipid reservoirs (at fixed chemical
potentials)
\cite{raucher99}.
A constant pressure in closed vesicles is also maintained via osmosis.
Therefore, for our study we choose the ensemble in which the surface
tension $\sigma$ and the inside pressure $p$ (relative to the outside)
are fixed rather than the surface area of the membrane $A$ or the
volume of the vesicle $V$.

With the bending term included the free energy of the membrane can be
written as
\cite{helfrich}
\be
{\cal F}=\int\frac{\kappa}{2}(2H)^2 \,{\rm d}A +\sigma A -pV -fL
\;,
\label{eq_F}
\ee
where
$\kappa$ is the bending rigidity and
$H$ is the mean curvature of the membrane.
The membrane is pulled in the $Z$ direction with a point force $f$, and
the end-to-end distance of the membrane in this direction is denoted
by~$L$.

For a tube of length $L$ and radius $R$ the free energy (at $p=0$) can
be written as
${\cal F}_{\rm tube}=[\kappa/(2R^2)+\sigma]2\pi RL - fL$.
To minimize ${\cal F}_{\rm tube}$ the surface tension acts to reduce
the radius, while the bending rigidity works against this. The balance
between the two sets the equilibrium radius $R_0$ and force $f_0$,
which can be calculated by taking
$\partial{\cal F}_{\rm tube}/\partial R=0$ and
$\partial{\cal F}_{\rm tube}/\partial L=0$:
\be
R_0=\sqrt{\frac{\kappa}{2\sigma}}
\; , \qquad
f_0=2\pi\sqrt{2\sigma\kappa}
\; .
\label{eq_R0_f0}
\ee
For typical values of
$\kappa\approx40$~pNnm and
$\sigma\approx0.05$~pN/nm one finds
$R_0\approx20$~nm and
$f_0\approx12.6$~pN.

\begin{figure}[!t]
\centerline{\includegraphics[scale=1]{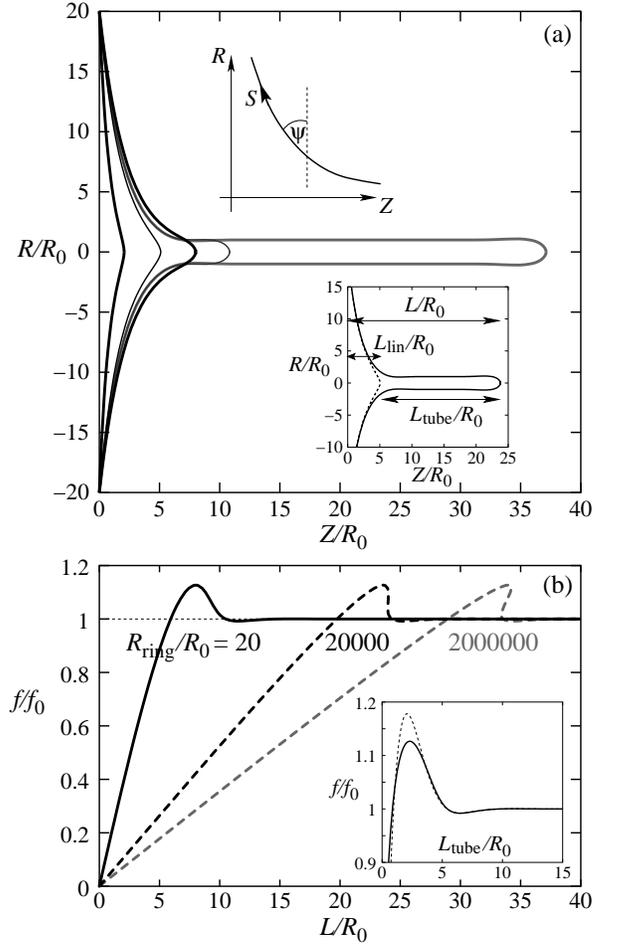}}
\caption{
(a) The shape of an emerging tube at various lengths. The upper inset
shows the parametrization of the surface, and the lower one illustrates
the definition of $L_{\rm tube}$, which is the size of the deviation of
the shape from the linear approximation (dashed line). (b) Force vs.\
length curves for three different ring sizes. The inset shows the
$f$-$L_{\rm tube}$ curve (solid line) and its asymptotic fit by Eq.\
(\ref{eq_fosc}) (dashed line).
}\label{f_1}\end{figure}

For simplicity we consider only axisymmetric surfaces with the $Z$ axis
being the symmetry axis (Fig.~\ref{f_1}a upper inset). Such surfaces
can be parametrized by the angle $\psi(S)$, where $S$ is the arclength
along the contour. The coordinates $R(S)$ and $Z(S)$ depend on
$\psi(S)$ through
\be
\dot R=\cos\psi \qquad \mbox{and} \qquad \dot Z=-\sin\psi
\; ,
\label{eq_RZ}
\ee
and the mean curvature can be expressed as
\be
2H=\dot\psi+(\sin\psi)/R
\; .
\label{eq_H}
\ee

For axisymmetric surfaces one can derive the so-called general shape
equation from the free energy (\ref{eq_F}) by variational methods
\cite{axisymmetric,firstint}:
\bea
\dddot\psi&=&-\frac{1}{2}\dot\psi^3 -\frac{2\cos\psi}{R}\ddot\psi
 +\frac{3\sin\psi}{2R}\dot\psi^2
 +\frac{3\cos^2\psi-1}{2R^2}\dot\psi
\nonumber \\ &&
 +\bar\sigma\dot\psi
 -\frac{\cos^2\psi+1}{2R^3}\sin\psi
 +\frac{\bar\sigma}{R}\sin\psi
 -\bar p
\; ,
\label{eq_shape}
\eea
where 
$\bar\sigma=\sigma/\kappa=1/(2R_0^2)$ and
$\bar p=p/\kappa$.
Taking the first integral of this equation leads to
\cite{firstint}
\bea
\ddot\psi\cos\psi&=&-\frac{1}{2}\dot\psi^2\sin\psi
 -\frac{\cos^2\psi}{R}\dot\psi
 +\frac{\cos^2\psi+1}{2R^2}\sin\psi
\nonumber \\ &&
 +\bar\sigma\sin\psi
 -\frac{1}{2}\bar p R -\frac{\tilde f}{R}
\; , \qquad
\label{eq_shape2}
\eea      
with the integration constant
$\tilde f=f/(2\pi\kappa)=f/(f_0 R_0)$.

For nearly flat membranes ($\psi\ll 1$), after the parameter change
$[\psi(S),R(S)]\to\psi(R)$, one can expand the shape
equation (\ref{eq_shape2}) in powers of $\psi$ and keep only the terms
up to linear order:
\be
R^2 \psi''+R\psi'-(R^2 \bar\sigma+1)\psi=-\tilde f R - \bar p R^3 /2
\; .
\label{eq_flat}
\ee
Because of the parameter change, the primes denote derivations with
respect to $R$ (which, in linear order of $\psi$, are identical to
derivations with respect to $S$). The general solution of this
differential equation is
\be
\psi(R)=\frac{\tilde f}{\bar\sigma}\frac{1}{R}
 +\frac{1}{2}\frac{\bar p}{\bar\sigma}R
 +c_1 I_1(R\sqrt{\bar\sigma}\,)
 +c_2 K_1(R\sqrt{\bar\sigma}\,)
\; , \;
\label{eq_psi}
\ee
where $I_i(x)$ and $K_i(x)$ are modified Bessel functions, and
$c_1$ and $c_2$ are integration constants.
$I_1(R\sqrt{\bar\sigma}\,)$ diverges exponentially for
$R\to\infty$. Because for a big vesicle we expect the shape to
converge to that of a sphere, $c_1$ must vanish. At $R=0$ the
divergence of the $1/R$ term must be canceled by the $K_1$ term,
leading to
$c_2=-\tilde f/\sqrt{\bar\sigma}$. Integrating $-\psi(R)$ with respect
to $R$ gives the shape of the membrane in this linear approximation:
\be
Z_{\rm lin}(R)=Z_0-\frac{2R_0 f}{f_0}
 \!\left[ \ln\!\left(\!\frac{R}{\sqrt{2}R_0}\right)\!
         +K_0\!\left(\!\frac{R}{\sqrt{2}R_0}\right)\! \right]\!
 -\frac{R^2}{2R_{\rm ves}},
\label{eq_Zlin}
\ee
where the integration constant $Z_0$ serves as a reference coordinate,
and we have expressed
$\sigma$, $\kappa$, and $p$ in terms of $R_0$, $f_0$, and vesicle
radius $R_{\rm ves}=2\sigma/p$. The last term is a trivial
contribution, describing a spherical vesicle under tension $\sigma$ and
pressure $p$. The second term, which is proportional to $f$, is the
linear response, and describes the deformation of the vesicle. The
quantity between the brackets converges to
$[\ln(2)-\gamma]$ for $R\to 0$, where $\gamma=0.577...$
is the Euler constant.
For $p=0$ and large $R$ the logarithmic term dominates, which
corresponds to a catenoid, the well known shape of a soap film in
cylindrical geometry under zero pressure.

Because the pressure makes only a trivial contribution, and has a
negligible effect on tube formation from big vesicles (plug
$pR\approx pR_0=2\sigma R_0/R_{\rm ves}\ll\sigma$ into the shape
equations), we neglect it from now on, and consider a piece of
(initially flat) membrane that spans a ring of radius $R_{\rm ring}$
located at $Z=0$.

For large deformations (or pulling forces) the linear
approximation breaks down, and we have to solve Eq.\ (\ref{eq_shape})
together with Eq.\ (\ref{eq_RZ}) numerically. Note that Eq.\
(\ref{eq_shape2}) could also be used, but it is numerically less
stable. We start solving the differential equations from the ring,
where we impose a zero curvature (or free hinge) boundary condition.
Thus, the four initial parameters at $Z=0$ are as follows:
 (i)
$R=R_{\rm ring}$;
 (ii)
$\psi=\arcsin[f/(2\pi\sigma R_{\rm ring})]-\varepsilon$, where the
small deviation $\varepsilon$ from the catenoid shape is chosen
(with a shooting and matching technique)
such that the contour line reaches the $Z$ axis;
 (iii)
$\dot\psi=-(\sin\psi)/R$ ensures zero mean curvature [see
Eq.\ (\ref{eq_H})];
 (iv) finally,
$\ddot\psi$ is determined from Eq.\ (\ref{eq_shape2}).

The results of the numerical solution can be seen in Fig.~\ref{f_1}.
The main part of Fig.~\ref{f_1}a shows the shape (contour line) of the
membrane for $R_{\rm ring}=20R_0$ and different values of $L$. For
small deformations (left line) Eq.\ (\ref{eq_Zlin}) gives a very good
approximation of the shape. In this linear regime the size of the
deformation is approximately
$L_{\rm lin}=Z_{\rm lin}(0)-Z_{\rm lin}(R_{\rm ring})$.

For larger deformations a tube emerges in the middle, and the linear
approximation fails. However, far from the tubular part, where $\psi$
is small, the approximation is still valid (Fig.~\ref{f_1}a lower
inset). Thus, it is convenient to define the tube as the piece between
$L_{\rm lin}$ and $L$, and the base as the rest of the membrane between
0 and $L_{\rm lin}$. This way, the dependence of the size of the total
deformation $L$ on the ring radius $R_{\rm ring}$ is absorbed in the
size of the base $L_{\rm lin}$, and the length of the tube $L_{\rm
tube}=L-L_{\rm lin}$ ({i.e.}, the deviation from the linear
approximation) becomes independent of $R_{\rm ring}$.

With this definition it is enough to determine the $f$-$L$ curve for
one particular ring size ({e.g.}, $20R_0$), from which the
universal (ring size independent) $f$-$L_{\rm tube}$ curve can be
calculated (Fig.~\ref{f_1}b inset). Because only the $f$-$L$ curves
have real physical meaning, one can easily calculate them from the
$f$-$L_{\rm tube}$ curve for any ring radius ($\gg R_0$) by simply
adding $L_{\rm lin}(f,R_{\rm ring})$ to
$L_{\rm tube}$. 
The most intriguing feature of the $f$-$L$ curves (shown in
Fig.~\ref{f_1}b for three different values of $R_{\rm ring}$)
is their non-monotonicity. The force first grows linearly,
in accordance with the linear approximation (\ref{eq_Zlin}), and
converges to $f_0$ for large $L$. But in between it overshoots by $\sim
0.13f_0$, and then oscillates about $f_0$ with an exponentially
decaying amplitude. This oscillation results in infinitely many
intervals with negative slopes, which, in the $f$-ensemble (where $f$
is the control parameter rather than $L$), are mechanically unstable
and represent an infinite series of {\it first order shape transitions}
at $f_0$. Because $L_{\rm lin}$ is a monotonically increasing function
of both $f$ and $R_{\rm ring}$, the main peak becomes an overhang for
large rings ($>20000R_0$), leading to a first order transition even in
the $L$-ensemble.

\begin{figure}[!t]
\centerline{\includegraphics[scale=1]{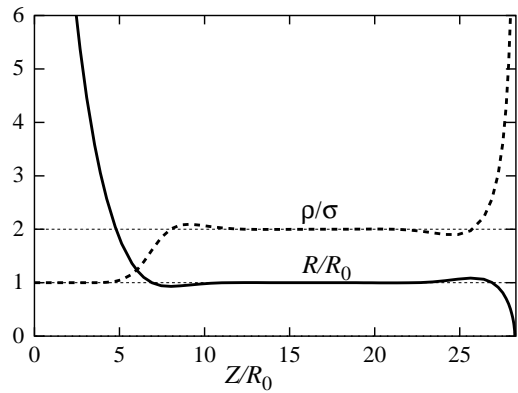}}
\caption{
The shape of a tube magnified in the radial direction (solid line), and
the surface energy density (dashed line).
}\label{f_2}\end{figure}

For a nearly cylindrical section, like the tubular part of the
membrane, the shape equations can be expanded in powers of
$U(Z)=R(Z)-R_0$.
After changing parameters $[\psi(S),Z(S)]\to\psi(Z)$, the expansion of
Eq.\ (\ref{eq_shape}) up to first order in $U(Z)$ and for $p=0$ reduces
to
\cite{bozic01}
\be
R_0^4 U''''=-U
\; .
\label{eq_Uexp}    
\ee
Here the primes denote derivations with respect to $Z$.
The solution of this equation is the sum of two exponentially decaying
oscillations from the two ends ($L_{\rm lin}$ and $L$) of the tube:
\bea
\frac{U(Z)}{R_0}=&&
 a_1\exp\!\left(\!-\frac{Z-L_{\rm lin}}{\sqrt{2}R_0}           \right)\!
    \cos\!\left(\! \frac{Z-L_{\rm lin}}{\sqrt{2}R_0}+\alpha_1\!\right)\!
\nonumber \\
 +&&
 a_2\exp\!\left(\!-\frac{L-Z          }{\sqrt{2}R_0}           \right)\!
    \cos\!\left(\! \frac{L-Z          }{\sqrt{2}R_0}+\alpha_2\!\right)\!
\; , \quad
\label{eq_Uosc}
\eea
where the integration constants converge to
$a_1\approx 0.746$,
$a_2\approx 0.726$,
$\alpha_1\approx 0.347$, and
$\alpha_2\approx 3.691$
(determined by numerical fitting) as the length of the tube increases.
These oscillations can be seen in Fig.~\ref{f_2} as overshootings from
both the base and the tip of the tube. They can be understood
intuitively by noticing that the smaller (larger) mean curvature near
the base (tip) makes the effect of the bending rigidity on the tube
radius less (more) pronounced, leading to a smaller (larger) $R$.

The expansion of Eq.\ (\ref{eq_shape2}) up to the first non-vanishing
order in $U(Z)$ gives us the deviation of the force:
\be
\frac{f-f_0}{f_0}=R_0^2U'U'''-\frac{1}{2}R_0^2 U''^2
 +\frac{1}{2}\frac{1}{R_0^2}U^2
\; .
\label{eq_fexp}
\ee
Plugging Eq.\ (\ref{eq_Uosc}) into Eq.\ (\ref{eq_fexp}) yields
\be
\frac{f-f_0}{f_0}=2a_1 a_2
    \exp\!\left(\!-\frac{L_{\rm tube}}{\sqrt{2}R_0}           \right)\!
    \cos\!\left(\! \frac{L_{\rm tube}}{\sqrt{2}R_0}+\alpha_1+\alpha_2\!\right)\!
\; ,
\label{eq_fosc}
\ee
which is an exponentially decaying oscillation as a function of
$L_{\rm tube}$. This explains the observed non-monotonic behavior of
the $f$-$L$ curves, and gives a very good fit to the $f$-$L_{\rm tube}$
curve (Fig.~\ref{f_1}b inset) with the same values of $a_i$ and
$\alpha_i$ as determined above for the shape oscillations. In a
different ensemble, Heinrich {\it et al.}\
\cite{heinrich99}
also observed similar oscillations for axially strained vesicles.

The surface energy density $\rho=(\kappa/2)(2H)^2+\sigma$ of the
membrane is plotted in Fig.~\ref{f_2}. It grows from $\sigma$ to
$2\sigma$ as we enter the tube from the base, and diverges at the tip.
Because at the tip the angle $\psi$ is small again, Eq.\ (\ref{eq_psi})
can be used to describe the surface there, leading to a logarithmic
divergence of the mean curvature
\cite{podgornik95,bozic01}.

In experiments, bundles of tubes can often be observed
\cite{roux02}.
To study the interaction between two tubes, let us start with a planar
membrane that spans a ring and is pulled perpendicularly by two point
forces ($f_1$ and $f_2$) at a distance $d$. If $d$ is large enough
($R_0\ll d \ll R_{\rm ring}$), both protrusions (except for the
vicinity of the points of pulling) can be described by the leading
logarithmic term of the linear approximation (\ref{eq_Zlin}). Their
superposition can be used to calculate the $d$-dependence of the free
energy of the membrane. Note that $fL$ in Eq.\ (\ref{eq_F}) has to be
replaced by $f_1 L_1+f_2 L_2$, where $L_1$ and $L_2$ are the sizes of
the deformations. Straightforward calculation results in an
attractive potential
${\cal F}(d)={\rm const} + 2R_0 \ln(d) f_1 f_2 /f_0$
between the two deformations. The same attraction has been found
between membrane-bound adhesion molecules
\cite{bruinsma98}.
For the attraction between two tubes
replace both $f_1$ and $f_2$ by $f_0$.

\begin{figure}[!t]
\centerline{\includegraphics[scale=1]{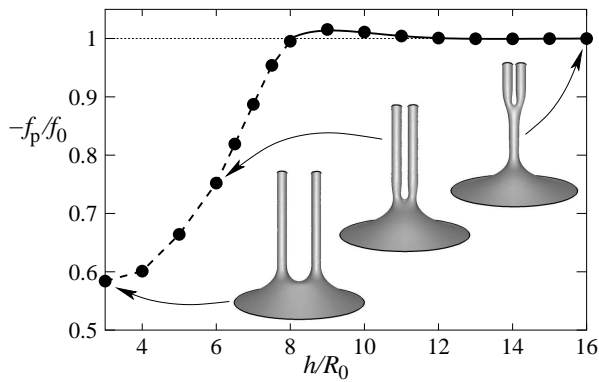}}
\caption{
The pinning force $f_{\rm p}$ as a function of $h$. It goes from
$-f_0/2$ (because if two tubes are far apart their attraction can be
eliminated by a $-f_0/2$ point force applied halfway) to
$-f_0$ (because for large $h$ the junction is pulled by two tubes
upward but by only one tube downward). The solid line is a fit by an
exponentially decaying oscillation.
}\label{f_3}\end{figure}

To see what happens when two tubes get close, we performed numerical
energy minimization with the {\sc Surface Evolver} program
\cite{brakke92}.
We defined the initial topology as a piece of membrane spanning a ring
of radius $12R_0$ at $Z=0$, with two cylindrical protrusions in the
positive $Z$ direction. We applied zero curvature boundary condition at
the ring and reflecting boundary condition at $Z=24R_0$
(Fig.~\ref{f_3}). To make sure that the two tubes do not coalesce we
placed a very narrow ($\ll R_0$) cylindrical obstacle between the
tubes, perpendicularly to them, and at a distance $h$ from the ring
(not shown in Fig.~\ref{f_3}). The obstacle simply pinned down the
middle of the membrane at $Z=h$. We then measured the pinning force
$f_{\rm p}$ (exerted by the obstacle on the membrane) as a function of
$h$. Fig.~\ref{f_3} shows that $f_{\rm p}$ is always negative ({\it
i.e.}, pushes the membrane), meaning that there is no energy barrier
against coalescence, it occurs smoothly. This is consistent with the
experiments of Evans {\it et al.}\
\cite{evans96}.

Tubes can be detached via fission and then transported individually in
the cell as long prolate-like vesicles (keeping their original radius
$R_0$). Because of volume and area constraints their energy contains
only the bending term: 
${\cal F}=\int(\kappa/2)(2H)^2 \,{\rm d}A$.
Although with such a high area-to-volume ratio the energetically most
favorable shape is the stomatocyte, these prolates are metastable
\cite{seifert91}.
They are also very flexible. If we consider a prolate as a rod and bend
it with a curvature $C$ ($\ll 1/R_0$), its energy, for symmetry
reasons, increases quadratically with $C$. This energy increase can be
written as
$(\kappa/2) (\lambda C^2) 2\pi R_0 L$,
where $L$ is the length of the prolate, and the factor
$\lambda=\left<\sin^2\phi\right>=1/2$
indicates that during the integration of the energy around a
cross-section (parametrized by $\phi$), only the out-of-plane component
of the bending counts. Thus, the bending stiffness of a prolate-like
vesicle is
$\kappa\pi R_0$,
and its persistence length
$\kappa\pi R_0/(k_{\rm B}T)$
is in the order of a few hundred nanometers.

Although these prolates are thicker than the microtubules, their
persistence length is much closer to that of a DNA. This is because
their wall is a two-dimensional fluid. Due to this fluidity they do not
even resist twisting, which makes them ideal semiflexible filaments.

We close this Letter by discussing some of the biological consequences of
our theoretical results.
The major biological relevance of the first force peak of the $f$-$L$
curves is that in order to form a tube, the motor proteins must be able
to provide a force that is 13\% larger than what is needed to pull a
long tube. At a pulling force of $f=f_0$ the peak corresponds to a
$2.1\kappa\approx21k_{\rm B}T$ high energy barrier, which is
practically insurmountable for such a big and slow object as a growing
tube. Thus, tube formation works on an all-or-nothing basis: motors can
pull out tubes only if they are strong enough to overcome the major
force peak.

Although at the tip of the tubes the size of the lipids ($\sim 0.5$~nm)
represents a natural cutoff length-scale for the divergence of the
energy density, it goes up to tens of $\sigma$. In terms of membrane
rupture, the energy density acts as an effective surface tension,
meaning that the most likely place for rupture to occur is the tip of
the tubes. So if biological systems want to avoid rupture, they either
have to protect the tips or distribute the pulling forces at larger
areas. That could be achieved, {e.g.}, by utilizing cap proteins or
lipid rafts.

We have shown that without external pinning tubes coalesce smoothly.
Thus, to explain bundle formation, other physical effects
that could prevent the coalescence of tubes
({e.g.}\ adhesion between the tubes and the cytoskeleton) must be
taken into account.

\begin{acknowledgments}
We thank
A. Roux, G. Cappello, P. Bassereau, and B. Goud
for their helpful discussions.
This work was supported by the
EU Marie Curie Fellowship HPMF-CT-1999-00124.
\end{acknowledgments}

\vspace{-10pt}							

\end{document}